# PHOTONUCLEAR REACTIONS CROSS-SECTIONS AT ENERGIES UP TO 100 MeV FOR DIFFERENT EXPERIMENTAL SETUPS

*O.S. Deiev, I.S. Timchenko, S.M. Olejnik, S.M. Potin, V.A. Kushnir,*
*V.V. Mytrochenko, S.A. Perezhogin, V.A. Bocharov, B.I. Shramenko*
*National Science Center "Kharkov Institute of Physics and Technology",*
*1 Akademichna St., Kharkiv, 61108, Ukraine*
*e-mail: timchenko@kipt.kharkov.ua*

In experiments on the electron linac LUE-40 of RDC "Accelerator" NSC KIPT, the flux-averaged cross-sections $\langle\sigma(E_{\gamma max})\rangle_{exp}$ of photonuclear reactions $^{100}$Mo$(\gamma,n)^{99}$Mo, $^{27}$Al$(\gamma,x)^{24}$Na, $^{93}$Nb$(\gamma,n)^{92m}$Nb, $^{93}$Nb$(\gamma,3n)^{90}$Nb, and $^{181}$Ta$(\gamma,n)^{180g}$Ta were measured using the γ-activation technique. The theoretical flux-average cross-sections $\langle\sigma(E_{\gamma max})\rangle_{th}$ were computed using the partial cross-section $\sigma(E)$ values from the TALYS1.9-1.95 codes and bremsstrahlung γ-flux calculated using GEANT4.9.2.

Two different experimental setups were used in the experiments: an aluminum electron absorber and a deflecting magnet to clean the bremsstrahlung γ-flux from electrons. A comparison of the flux-average cross-sections measured for two experimental setups was performed. The possibility of using the reactions $^{100}$Mo$(\gamma,n)^{99}$Mo, $^{27}$Al$(\gamma,x)^{24}$Na, $^{93}$Nb$(\gamma,n)^{92m}$Nb, $^{93}$Nb$(\gamma,3n)^{90}$Nb, and $^{181}$Ta$(\gamma,n)^{180g}$Ta as monitors of the bremsstrahlung γ-flux for the energy range 30–100 MeV was investigated.

## INTRODUCTION

Currently, experimental studies of many-particle photonuclear reactions in the energy range above the giant dipole resonance (GDR) are being actively carried out, for example, works [1-5].

Study of photofission of nuclei in the energy range above the GDR up to the pion production threshold ($E_{th} \approx 145$ MeV) is of particular interest because there is a change in the mechanism of interaction of photons with nuclei in this energy region. Therefore it is possible to obtain fundamental data on two mechanisms of photofission of the nucleus, namely, due to GDR excitation and quasi-deuteron photoabsorption [6]. To date, various theoretical models, for example, [7,8], as well as the modern nuclear reaction codes [9-12], have been developed to describe photonuclear reactions at their cross-section calculations. Both the theoretical models and the calculation codes need new data to verify calculations for multiparticle reactions in a wide range of atomic masses and energies.

In practical applications, knowledge of the exact parameters of multiparticle photonuclear reactions is important, for example, to estimate the neutron yield in nuclear installations based on accelerator-driven subcritical systems (ADS). The ADS can be used for the utilization of radioactive waste of nuclear energy plants (transmutation of minor actinides [13] and burning long-lived fission fragments). The ADS are also potential sources of electricity generation [14,15].

The multiparticle photonuclear experiments require the use of high-intensity gamma-ray fluxes. Present, either quasimonoenergetic annihilation photons or high-energy electron bremsstrahlung are used to determine cross-sections of partial reactions. In the first case, the absolute values of the cross-sections are measured in experiments, and in the second case, information is obtained on the flux-average or integral characteristics of investigated reactions.

To obtain cross-sections of the reaction, it is necessary to know the exact value of the flux of bremsstrahlung quanta on the target, which is usually calculated using the GEANT4 code. Some factors of an experiment can make an error: geometric (displacement of the target center relative to the beam axis), inaccuracy of the irradiation dose and beam current at longtime exposures, beam profile, etc.

For the control of the irradiation parameters, in the experiment can be used absolute measurements of bremsstrahlung flux with a quantometer. But a simpler and more reliable way is using targets-monitors with good know cross-sections and which are in the same conditions as the studied target [3,16,17]. The possibilty of the monitoring of γ-ray flux was studied, for example, in the works [18,19] using several reactions for a wide range of energies 300 – 1000 MeV.

To select the monitor, it needs to be guided by several criteria:

- The absolute cross-section, $\sigma(E)$, of the reaction chosen must be known with sufficient precision.
- The product of the reaction must be easily detectable by standard spectroscopy.
- The monitor reaction should yield a radioactive product with a half-life longer than the longest irradiation time to be monitored.
- The cost of foils to be activated should not be very expensive.

These requirements limit the choice of the monitor's target material.

To date, for example, to monitor the flux of quanta, the following reactions $^{197}$Au$(\gamma,n)^{196}$Au [20], $^{65}$Cu$(\gamma,n)^{64}$Cu [5,21], $^{100}$Mo$(\gamma,n)^{99}$Mo [22-24] are used. These reactions have reliable experimental cross-section data in the literature [25-27]. The high-threshold reaction $^{27}$Al$(\gamma,x)^{24}$Na, for which the data are presented as a cross-section per equivalent photon [19,28], is also used for gamma-flux monitoring [4].

The flux of bremsstrahlung quanta, which are generated when electrons pass through targets-converters, must be cleaned from impurity of the electron component. It is necessary to do since they make an additional contribution to the yield of studied reactions. Usually, this contribution can be calculated, although its probability is strongly suppressed by the mechanism of interaction of electrons through virtual photons. The constant of this interaction is proportional to the fine structure constant which is about 1/137. However, for high threshold reactions, the contribution of electrons may dominate. Thus, it was shown in the case of thin Ta-converter and without Al-absorber [29] on the example of $^{93}$Nb($\gamma$,$x$n)$^{93-x}$Nb reactions, where $x$ is neutron multiplicity, that for the case of $x$ = 5 the yield of the reaction under the action of gamma quanta becomes less than that from the contribution of electrons.

Thus, in this paper, we present the results of measurements at the electron linac LUE-40 of RDC "Accelerator" NSC KIPT for photonuclear reactions $^{100}$Mo($\gamma$,n)$^{99}$Mo, $^{27}$Al($\gamma$,$x$)$^{24}$Na, $^{93}$Nb($\gamma$,n)$^{92m}$Nb, $^{93}$Nb($\gamma$,3n)$^{90}$Nb, and $^{181}$Ta($\gamma$,n)$^{180g}$Ta in the bremsstrahlung end-point energy range $E_{\gamma max}$ = 35-95 MeV. These results were obtained in two different experimental approaches (Setup 1 and Setup 2) for cleaning a beam of bremsstrahlung gamma flux from impurity of the electronic component. A comparison of the experimental flux-average cross-sections measured for two Setups was performed.

The possibility of using the reactions $^{100}$Mo($\gamma$,n)$^{99}$Mo, $^{27}$Al($\gamma$,$x$)$^{24}$Na, $^{93}$Nb($\gamma$,n)$^{92m}$Nb, $^{93}$Nb($\gamma$,3n)$^{90}$Nb, and $^{181}$Ta($\gamma$,n)$^{180g}$Ta for monitoring the bremsstrahlung flux of gamma quanta in the end-point energy range of 30-100 MeV is considered.

## 1. EXPERIMENTAL PROCEDURE

In this paper, we consider two different schemes for performing studies of photonuclear reactions: in one of which an aluminum absorber is used to clean the bremsstrahlung flux from electrons, and in the second one, a deflecting magnet is used. The experimental Setups are schematically shown in Fig. 1.

### 1a. EXPERIMENTAL SETUP 1

In the first experimental scheme, a converter, passing through which electrons interact with atoms of the substance and emit bremsstrahlung radiation, is used to obtain a flux of gamma-quanta. To clean the bremsstrahlung $\gamma$-quanta flux from electrons, passed through the converter, a massive electron absorber consisting of light material (usually Al) [30,31] is installed. This makes it possible to obtain an almost "pure" gamma-ray flux on the target.

The advantages of this method are mainly in the simplicity of the approach. This guarantees a slight radiation and heat load of the target and, for example, on the components of the pneumatic tube transport, which is used for target delivery.

The disadvantages of this scheme are the distortion of the shape of the bremsstrahlung spectrum, and the additional generation of photoneutrons, which also contribute to studied reaction yield. In addition, in such a scheme of the experiment, modelling the bremsstrahlung flux in the GEANT4 code is complicated.

### 1b. EXPERIMENTAL SETUP 2

The second scheme of the experiment implements a deflecting magnet to divert the electrons that have passed through the converter [31] that allowing obtaining a "pure" beam of bremsstrahlung gamma quanta at the target [32]. This assumes the use of a thin Ta-converter to minimize the spread of the electron beam. When using thin converter, the shape of the radiation spectrum can be described by a known analytical formula. At the same time, the influence of neutrons is minimized.

Thus, the study of photonuclear reactions with the "pure" gamma beam using this scheme has the following advantages:
- Deflected electrons will not create a radiation load on the target and in the pneumatic system. Gamma rays will not scatter in the absorber.
- The bremsstrahlung spectrum can be described by an analytical formula of the form ~ 1/$E_\gamma$. There are advantages in obtaining cross-sections for photonuclear reactions $\sigma(E)$ by the photon difference method or the regularization method, associated with a better shape of the "difference peak".
- Due to the absence of the absorber, it is possible to significantly decrease the volume of calculations of bremsstrahlung quantum flux with the computer codes. Experimental and calculated errors are significantly reduced.
- The contribution of neutrons to the reaction yields can be neglected.

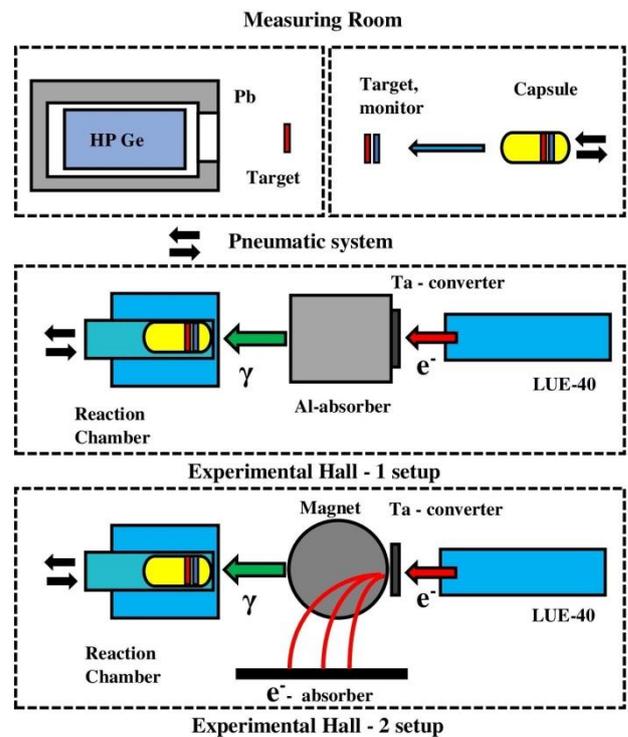

*Fig. 1. Experimental schematic diagram. Above – the measuring room, below – the experimental hall of the LUE-40 accelerator: Setup 1 – with Al-absorber, Setup 2 – with a deflecting magnet. Between the measuring room and the experimental hall, there is a pneumatic mail system.*

The GEANT4.9.2 code [33] was used to calculate the neutron yield from the target and structural elements of the facility under experimental conditions. PhysList *QGSP_BIC_HP* makes it possible to calculate the neutron yield due to photonuclear reactions from targets of different thicknesses and elements. Using the calculation method described in [34] we have estimated the total neutron yield in the full angle $4\pi$ for different target-converter thicknesses. For example, at electron energy of $E_e$ = 100 MeV for Ta-converters 0.1 mm and 1 mm thick, the neutron yields are 0.0023 and 0.02 n/kW/s, respectively. Whereas for the combination of Ta 1 mm and Al 100 mm, the electron yield sharply increases to 0.33 n/kW/s.

Thus, the use of a scheme with a thin converter and without Al-absorber leads to a decrease in the total number of neutrons by at least 100 times. This removes the question of the neutrons contribution to the yield of a photonuclear reaction.

### 1c. EXPERIMENTAL PROCEDURE

The experiments were performed on the bremsstrahlung gamma-beam from the electron linear accelerator LUE-40 RDC "Accelerator" NSC KIPT using the method of induced γ-activity of the final product nucleus of the reaction. The experimental procedure is described in detail in [1,2].

The studies on the linac LUE-40 [35] performed at end-point energy range 35–95 MeV (this coincides with the energy of the electron beam $E_e$). The average current of the beam is $I_e \approx$ 3-4 μA. The electron energy spectrum width at the full width at half maximum (FWHM) makes $\Delta E_e/E_e \sim$ 1%.

In the case of experimental Setup 1 – the bremsstrahlung gamma radiation was generated by passing a pulsed electron beam at end-point energy range 35–95 MeV through a tantalum metal plate, 1.05 mm in thickness. The Ta-converter was fixed on the aluminum cylinder, 100 mm in diameter and 150 mm in thickness. For Setup 2 – the bremsstrahlung radiation was generated by passing an electron beam at end-point energies $E_{\gamma max}$ = 42.4; 60.4 and 80.5 MeV through the Ta foil, 0.1 mm in thickness and without Al-absorber.

In experiments for both Setup 1 and 2, the natural Mo, Al, Ta and Nb targets, which represented thin discs with diameters 8 mm, were used. The target masses were: $m_{Mo} \approx$ 60 mg, $m_{Al} \approx$ 135 mg, $m_{Nb} \approx$ 80 mg, and $m_{Ta} \approx$ 43 mg.

For transporting the capsule with the sample to the place of irradiation and back for induced activity registration, a pneumatic tube transfer system was used. On delivery of the irradiated targets to the measuring room, the samples are extracted from the aluminium capsule and were transferred one by one to the detector for the measurements. Taking into account the time of target delivery and extraction from the capsule, the cooling time for the sample under study took no more than 3 minutes.

The γ-quanta of the reaction products were detected using a Canberra GC-2018 semiconductor HPGe detector with the relative detection efficiency of 20%. The resolution FWHM is 1.8 keV for energy $E_\gamma$ = 1332 keV and is 0.8 keV for $E_\gamma$ = 122 keV. The dead time for γ-quanta detection varied between 0.1 - 5%. The absolute detection efficiency for γ-quanta of different energies was obtained using a standard set of γ-quanta sources: $^{241}$Am, $^{133}$Ba, $^{60}$Co, $^{137}$Cs, $^{22}$Na, $^{152}$Eu.

The bremsstrahlung flux was monitored by the yield of the $^{100}$Mo(γ,n)$^{99}$Mo, $^{27}$Al(γ,x)$^{24}$Na, $^{93}$Nb(γ,n)$^{92m}$Nb, $^{93}$Nb(γ,3n)$^{90}$Nb, and $^{181}$Ta(γ,n)$^{180g}$Ta reactions. In this work, we consider a target from $^{nat}$Mo as a main universal monitor. Nuclear spectroscopic data of the radionuclides and reactions presented in Table 1 [36].

Table 1. Nuclear spectroscopic data of the reaction products from ref. [36]

| Reaction | $E_{th}$, MeV | $T_{1/2}$ | $E_\gamma$, keV ($I_\gamma$, %) |
|---|---|---|---|
| $^{100}$Mo(γ,n)$^{99}$Mo | 8.29 | 65.94 h | 739.50 (12.13) |
| $^{27}$Al(γ,2pn)$^{24}$Na | 31.43 | 14.958 h | 1368.633 (100) |
| $^{93}$Nb(γ,n)$^{92m}$Nb | 8.45 | 10.15 d | 934.46 (99) |
| $^{93}$Nb(γ,3n)$^{90}$Nb | 28.77 | 14.60 h | 1129.22 (92.7) |
| $^{181}$Ta(γ,n)$^{180g}$Ta | 7.58 | 8.152 h | 103.557 (0.81) |

The electron bremsstrahlung spectra were calculated using the open-source software code GEANT4.9.2 [33]. The real geometry of the experiments was used in calculations also the space and energy distributions of the electron beam were taken into account.

The uncertainty of experimental values of the flux-average cross-sections $\langle \sigma(E_{\gamma max}) \rangle$ was determined as a quadratic sum of statistical and systematical errors. The statistical error of the observed γ-activity is mainly due to statistics calculation in the full absorption peak of the corresponding γ-line, which varies within 1 to 10%.

The systematic error is associated with the procedure of processing the measured data (the irradiation time $t_{irr}$, the electron current $I_e$; the detection efficiency ε, the value of the gamma-ray flux $N_\gamma$, the shape of the background substrate in the γ-spectrum of induced activity). There are also errors with the uncertainties in the tabular data for the half-life period $T_{1/2}$ and the absolute intensity $I_\gamma$. Description of these errors can be found in [2].

### 1d. CALCULATION FORMULAS

To calculate the theoretical value of the flux-average cross-section $\langle \sigma(E_{\gamma max}) \rangle_{th}$, the absolute cross-sections σ(E) for monoenergetic γ-quanta from the TALYS1.9-1.95 [10] codes were used with default parameters. The calculated cross-sections σ(E) were then averaged over the bremsstrahlung flux $W(E,E_{\gamma max})$, computed in GEANT4.9.2 code, in the energy range from the threshold $E_{th}$ of the corresponding reaction to the maximum energy of the bremsstrahlung γ-quanta spectrum $E_{\gamma max}$. As a result, the theoretical values of the bremsstrahlung flux-average cross-sections were obtained:

$$\langle \sigma(E_{\gamma max}) \rangle = \frac{\int_{E_{th}}^{E_{\gamma max}} \sigma(E) W(E, E_{\gamma max}) dE}{\int_{E_{th}}^{E_{\gamma max}} W(E, E_{\gamma max}) dE}. \quad (1)$$

The calculated in this way $\langle\sigma(E_{\gamma\max})\rangle$ values were compared with the experimental measured flux-average cross-sections, determined by the expression:

$$\langle\sigma(E_{\gamma\max})\rangle = \frac{\lambda\Delta A}{\varepsilon N_x I_\gamma \Phi(E_{\gamma\max})(1-e^{-\lambda t_{irr}})e^{-\lambda t_{cool}}(1-e^{-\lambda t_{meas}})}, \quad (2)$$

where $\Delta A$ is the number of counts of γ-quanta in the full absorption peak (for the γ-line of the investigated reaction), $\Phi(E_{\gamma\max}) = \int_{E_{th}}^{E_{\gamma\max}} W(E, E_{\gamma\max})dE$ is the bremsstrahlung quanta flux in the energy range from the reaction threshold $E_{th}$ up to $E_{\gamma\max}$, $N_x$ is the number of target atoms, $I_\gamma$ – the absolute intensity of the analyzed γ-quanta, $\varepsilon$ – the absolute detection efficiency for the analyzed γ-quanta energy, $\lambda$ is the decay constant ($\ln 2/T_{1/2}$), $t_{irr}$, $t_{cool}$ and $t_{meas}$ are the irradiation time, cooling time and measurement time, respectively.

From Eq. (1) and (2) it follows that the value of the flux-average cross-section $\langle\sigma(E_{\gamma\max})\rangle$ depends on the energy distribution of the bremsstrahlung flux and the value of the reaction threshold $E_{th}$. Due to the differences between these two important parameters, the cross-section values both $\langle\sigma(E_{\gamma\max})\rangle_{th}$ and $\langle\sigma(E_{\gamma\max})\rangle_{exp}$ may differ slightly for different Setup 1 and 2, as shown in work [2].

To account for a deviation of the GEANT4.9.2-computed bremsstrahlung γ-flux from the real flux falling on the target, the normalization (monitoring) factor $k_{monitor}$ must be calculated. For this, control measurements of reaction yield on a monitor target with a well-known cross-section are used. The value of monitoring factor $k_{monitor}$ is obtained from comparison the calculated and measured cross-sections as in [3]

$$k_{monitor} = \langle\sigma(E_{\gamma\max})\rangle_{th} / \langle\sigma(E_{\gamma\max})\rangle_{exp}. \quad (3)$$

Thus, we find the transition coefficients $k_{monitor}$ from calculated to experimental flux.

## 2. RESULTS AND DISCUSSIONS

When using the different experimental setups (in our case Setup 1 and Setup 2), the installation parameters, in particular the distance between an investigated target and a convertor, change significantly. The flux of bremsstrahlung photons, the method of removing electron component from bremsstrahlung γ-flux, the number of generated neutrons, etc. have also changed. It is important to show that the experimental cross-sections for photonuclear reactions for the two setups of facilities will coincide.

In the present work, new experimental data for photonuclear reactions $^{100}$Mo(γ,n)$^{99}$Mo, $^{27}$Al(γ,x)$^{24}$Na, $^{93}$Nb(γ,n)$^{92m}$Nb, $^{93}$Nb(γ,3n)$^{90}$Nb, and $^{181}$Ta(γ,n)$^{180g}$Ta are obtained using Setup 1 and 2 at end-point energy range up to 100 MeV.

### 2a. THE $^{100}$Mo(γ,n)$^{99}$Mo REACTION

The bremsstrahlung gamma flux monitoring against the $^{100}$Mo(γ,n)$^{99}$Mo reaction yield was performed by comparing the experimentally obtained average cross-section values with the computation data. To determine the experimental $\langle\sigma(E_{\gamma\max})\rangle_{exp}$ values by Eq. (2), we have used the $\Delta A$ activity value for the $E_\gamma = 739.50$ keV γ-line and the absolute intensity $I_\gamma = 12.13\%$ (see Table 1). The theoretical values of the average cross-section $\langle\sigma(E_{\gamma\max})\rangle_{th}$ were calculated by Eq. (1) using the cross-sections $\sigma(E)$ in the TALYS1.9-1.95 codes with the default options.

It is also a γ-line with $E_\gamma = 140.51$ keV, $I_\gamma = 89.43\%$ in a spectrum of induced gamma-activity of the $^{99}$Mo nucleus. Despite the high yield of the reaction by this γ-line, some disadvantages are not allowing it to be used for monitoring:
- the complex shape of the background under the total absorption peak,
- the energy $E_\gamma = 140.51$ keV is at the region of the peak of the detection efficiency curve of the detector; calibration has a maximum error,
- there is γ-line from niobium with a similar energy ($^{90}$Nb, $E_\gamma = 140.18$ keV, $I_\gamma = 66.8\%$, $T_{1/2} = 14.6$ h), which requires long-term cooling of the Mo target.

The cross-sections of the $^{100}$Mo(γ,n)$^{99}$Mo reaction were measured independently for two γ-lines $E_\gamma = 140.51$ and $739.50$ keV. The discrepancy between the experimental data was within the interval 0.5-2.8%.

The possibility of using the theoretical cross-sections $\sigma(E)$ from the TALYS1.9 code for the calculation was previously analyzed by us in [2]. The experimental cross-sections $\sigma(E)$ from [26] were compared with the calculations from TALYS1.9 with the default options. Then experimental cross-sections $\sigma(E)$ [26] were approximated by the analytical curve using the Lorenz formula. The average cross-sections $\langle\sigma(E_{\gamma\max})\rangle_{th}$ were calculated using both the $\sigma(E)$ values from the TALYS1.9 code and resulted fitting curve. As a result, it was shown that the difference between these average cross-sections $\langle\sigma(E_{\gamma\max})\rangle_{th}$ does not exceed 1.9% in the energy range $E_{\gamma\max} = 30$–100 MeV. These conclusions are also valid in the case of using a theoretical calculation based on the cross-section from the TALYS1.95 code. The monitoring procedure has been detailed in [2,16].

The normalization coefficients $k_{monitor}$ display the deviation of the calculated in the GEANT4.9.2 code bremsstrahlung flux from the real flux that fell on the target. The found values of $k_{monitor}$, which varied for different Setups and experiment conditions, were used to normalize the cross-sections of the other investigated photonuclear reactions. For example, Fig. 2 shows experimental $\langle\sigma(E_{\gamma\max})\rangle_{exp}$ and theoretical $\langle\sigma(E_{\gamma\max})\rangle_{th}$ flux-average cross-sections for the $^{100}$Mo(γ,n)$^{99}$Mo reaction obtained for Setup 1 in work [16].

The using of the $^{100}$Mo(γ,n)$^{99}$Mo reaction as the monitor for bremsstrahlung flux has some advantages. Firstly, the experimental cross-section $\sigma(E)$ [26] is large and coincides with the TALYS1.9-1.95 codes. Second, the half-life period is convenient for studying different reactions, and the reaction threshold allows using it from 10 MeV. Natural Mo contains several isotopes, which leads to a complex emission spectrum with a large number of γ-lines, some of which are composed of radiation from different residual nuclei. However, there are convenient lines for obtaining the reaction yield.

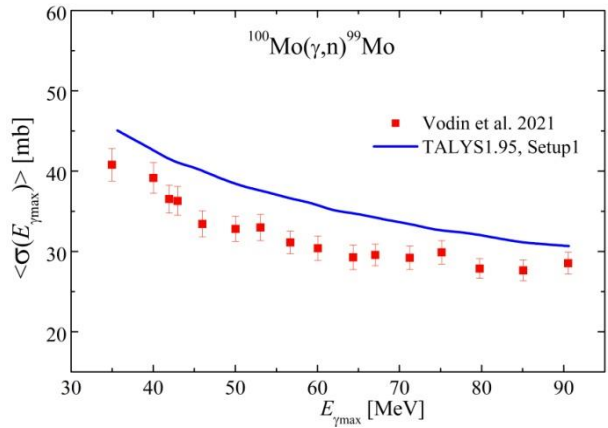

*Fig. 2. The flux-average cross-sections $\langle\sigma(E_{\gamma max})\rangle$ for the $^{100}Mo(\gamma,n)^{99}Mo$ reaction. Experimental data are points from work [16], theoretical calculation is blue line.*

### 2b. THE $^{27}Al(\gamma,x)^{24}Na$ REACTION

In addition to $^{nat}Mo$, the photonuclear reactions on $^{27}Al$, $^{93}Nb$, and $^{181}Ta$ targets were measured to monitor the bremsstrahlung flux. The use of various photonuclear reactions increases the reliability of the bremsstrahlung flux measuring for different energies and makes it possible to complement and compare the experimental values of the reaction cross-sections for two different setups.

For the $^{27}Al(\gamma,x)^{24}Na$ reaction, an additional control series of measurements was carried out for Setup 1 and the following average cross-sections were obtained: at $E_{\gamma max}$ = 54.85 MeV – $\langle\sigma(E_{\gamma max})\rangle$ = 0.219 ± 0.017 mb; $E_{\gamma max}$ = 85.35 MeV – $\langle\sigma(E_{\gamma max})\rangle$ = 0.211± 0.016 mb.

For the $^{27}Al(\gamma,x)^{24}Na$ reaction on Setup 2, experimental values of the average cross-sections were obtained: at $E_{\gamma max}$ = 42.4 MeV – $\langle\sigma(E_{\gamma max})\rangle$ = 0.120 ± 0.012 mb; $E_{\gamma max}$ = 60.4 MeV – 0.223 ± 0.021 mb, $E_{\gamma max}$ = 80.5 MeV – 0.218 ± 0.021 mb. As can be seen from Fig. 3, the new results coincide within the error with our previous results for Setup 1 [3] and from work [28]. There is a scatter in the experimental cross-sections. First of all, this is due to the small value of the $^{27}Al(\gamma,x)^{24}Na$ reaction cross-section, especially in the near-threshold region.

The calculated cross-sections for the $^{27}Al(\gamma,x)^{24}Na$ reaction from Talys1.95 code does not agree with the experimental data available in the literature [3,28]. As was previously shown in [3], the difference between them averages 2.4 times at energies of 50–95 MeV and differs greatly in the near-threshold region. This leads to the conclusion that to monitor the bremsstrahlung γ-flux by the yield of the $^{27}Al(\gamma,x)^{24}Na$ reaction, it is necessary to use in Eq. (3) instead theoretical $\langle\sigma(E_{\gamma max})\rangle_{th}$ values experimental data from [3] and [28]. As a result, the $k_{monitor}$ values, in this case, will be equal to the ratio of a well-known experimental cross-section to obtained one. For the monitoring procedure, it is necessary to approximate the data from [3,28] with a smooth curve using the least squares method.

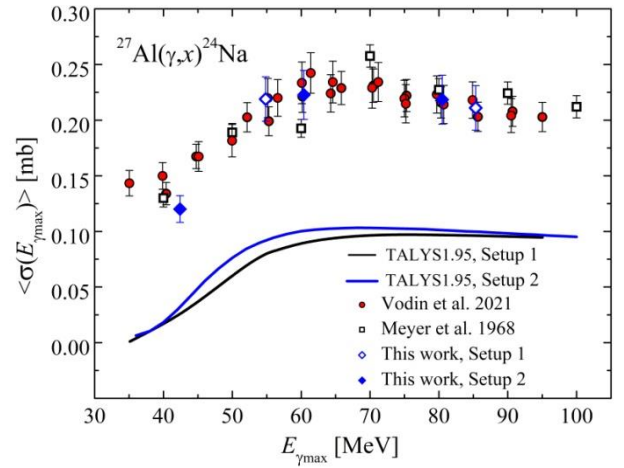

*Fig. 3. The flux-average cross-sections for the $^{27}Al(\gamma,x)^{24}Na$ reaction. Theoretical calculations are lines, experimental data: blue full and empty rhombus - this work, red filled circles - [3], squares - [28].*

The coincidence of $\langle\sigma(E_{\gamma max})\rangle_{exp}$ values from the present work and early known from [3,28] indicates that the $k_{monitor}$ for Mo and Al targets are equal. This is an important fact because thresholds of the $^{27}Al(\gamma,x)^{24}Na$ and $^{100}Mo(\gamma,n)^{99}Mo$ reactions are significantly different.

Aluminum is the most affordable material for flux monitoring. The emission spectrum of product-reaction $^{24}Na$ contains a small number of high-intensity lines. At the same time, $^{27}Al(\gamma,x)^{24}Na$ has a small cross-section and a high reaction threshold. Therefore, using this reaction as a monitor of gamma-flux is the most reliable in the bremsstrahlung end-point energy range $E_{\gamma max}$ > 45 MeV.

### 2c. THE $^{93}Nb(\gamma,n)^{92m}Nb$ AND $^{93}Nb(\gamma,3n)^{90}Nb$ REACTIONS

In this work for Setup 2, we measured the flux-average cross-sections for the $^{93}Nb(\gamma,n)^{92m}Nb$ and $^{93}Nb(\gamma,3n)^{90}Nb$ reactions.

For the $^{93}Nb(\gamma,n)^{92m}Nb$ reaction the following cross-section was obtained: $E_{\gamma max}$ = 42.4 MeV – $\langle\sigma(E_{\gamma max})\rangle$ = 19.2 ± 2.3 mb. To calculate the total cross-section for the $^{93}Nb(\gamma,n)^{92}Nb$ reaction we used a ratio value between metastable and ground states from TALYS1.9 code, which is equal to 0.551 at a wide energy range. As result the total cross-section $\langle\sigma(E_{\gamma max})\rangle$ = 34.9 ± 4.2 mb. The obtained cross-sections are in good agreement with the data from [2], and some discrepancies with [20] (see Fig. 4).

For the $^{93}Nb(\gamma,3n)^{90}Nb$ reaction, an additional control series of measurements was carried out for Setup 1 and the following flux-average cross-sections were obtained: $E_{\gamma max}$ = 50.2 MeV – $\langle\sigma(E_{\gamma max})\rangle$ = 3.11 ± 0.24 mb; $E_{\gamma max}$ = 79.95 MeV – $\langle\sigma(E_{\gamma max})\rangle$ = 2.56 ± 0.16 mb.

On Setup 2 for the $^{93}Nb(\gamma,3n)^{90}Nb$ reaction the flux-average cross-section were obtained at $E_{\gamma max}$ = 42.4 MeV – $\langle\sigma(E_{\gamma max})\rangle$ = 3.34 ± 0.21 mb. These are also in good agreement with the data [2], but there is some discrepancy with [20] (see Fig. 5).

As in case of the $^{27}Al(\gamma,x)^{24}Na$ reaction, the coincidence between the $\langle\sigma(E_{\gamma max})\rangle_{exp}$ values obtained in the present work and data from [2] for the

$^{93}$Nb(γ,n)$^{92m}$Nb and $^{93}$Nb(γ,3n)$^{90}$Nb reactions indicates that the values $k_{monitor}$ for Mo and Nb are equal.

Unfortunately, in the case of the $^{93}$Nb(γ,3n)$^{90}$Nb and $^{93}$Nb(γ,n)$^{92m}$Nb reactions, there are some discrepancies between the data [2] and [20]. And also, the theoretical calculations with the TALYS1.9 code differ from experimental data (see Fig. 4 and 5).

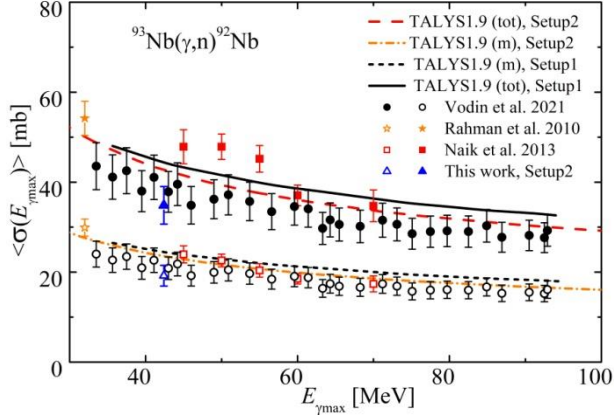

*Fig. 4. The flux-average cross-sections for the $^{93}$Nb(γ,n)$^{92m}$Nb reaction. The curves show the computations based on TALYS1.9 code. The experimental data triangles - this work, circles - [2], squares - [20], stars - [37]. Filled symbols are for the total average cross-sections and open ones are for the $^{93}$Nb(γ,n)$^{92m}$Nb average cross-sections.*

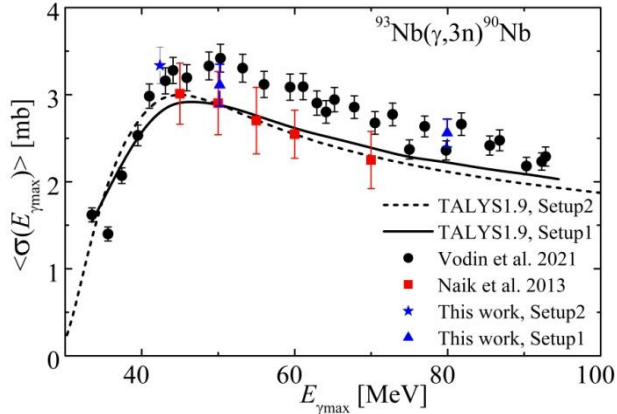

*Fig. 5. The flux-average cross-sections for the $^{93}$Nb(γ,3n)$^{90}$Nb reaction. The curves show the computations based on TALYS1.9. The experimental data: star and triangles - this work, circles - [2], squares - [20].*

Niobium is a monoisotopic element. The induced activity spectrum of irradiated Nb has a large number of γ-lines, but it is possible to find a convenient line to measure the reaction yield. The cross-section of both reactions is large, the half-life periods are suitable. The use of the $^{93}$Nb(γ,n)$^{92m}$Nb reaction as the monitor is possible from 10 MeV, in the case of the $^{93}$Nb(γ,3n)$^{90}$Nb reaction at energies from 40 MeV.

### 2d. THE $^{181}$Ta(γ,n)$^{180g}$Ta REACTION

For the $^{181}$Ta(γ,n)$^{180g}$Ta reaction on Setup 2, experimental flux-average cross-sections were obtained at: $E_{γmax} = 60.4$ MeV $- \langle σ(E_{γmax}) \rangle = 65.4 \pm 13.3$ mb, $E_{γmax} = 80.5$ MeV $- 64.6 \pm 13.2$ mb.

The total cross-section for the formation of the $^{180}$Ta nucleus was obtained using the ratio of the ground state cross-sections to the total $\langle σ(E_{γmax}) \rangle$, calculated in TALYS1.95 for the default parameters [1,38]. As a result, the experimental total flux-average cross-sections were obtained: $E_{γmax} = 60.4$ MeV $- \langle σ(E_{γmax}) \rangle = 71.4 \pm 14.6$ mb, $E_{γmax} = 80.5$ MeV $- 70.5 \pm 14.4$ mb.

As can be seen from Fig. 6, there is a coincidence between the experimental cross-sections within the error limits for both Setups. Unfortunately, the error of the obtained values is large, and it is associated with the magnitude of the unavoidable error of the tabular value the absolute intensity $I_γ = 0.0081 \pm 0.0016$ [36].

In the case of the $^{181}$Ta(γ,n)$^{180g}$Ta reaction, the calculated average cross-sections $\langle σ(E_{γmax}) \rangle_{th}$ are noticeably different for the two Setups. Nevertheless, the ratio of calculated values to experimental ones is approximately the same.

The $^{181}$Ta(γ,n)$^{180g}$Ta reaction as a monitor for bremsstrahlung flux is some flaws. Irradiated Ta has a complex emission spectrum with a large number of γ-lines. The γ-line associated with decay of $^{180}$Ta (103.557 keV) have a large tabular error of the absolute intensity $I_γ$. The reaction cross-section $σ(E)$ known from the literature [39,40] are differ from each other, and the experimental $\langle σ(E_{γmax}) \rangle$ [1,38] are somewhat lower than the calculation in TALYS1.95.

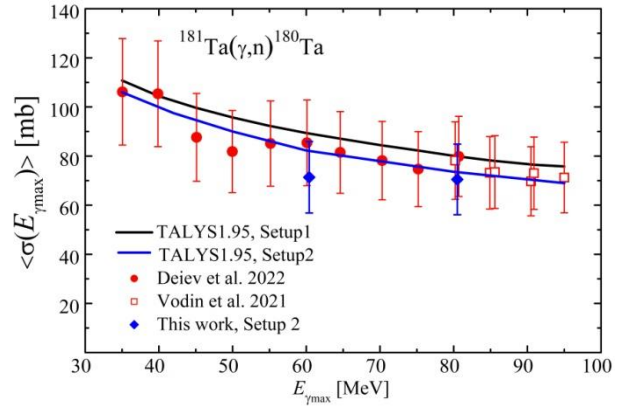

*Fig.6. Total average cross-sections $\langle σ(E_{γmax}) \rangle$ for the reactions $^{181}$Ta(γ,n)$^{180}$Ta. Full circles are from [1], empty squares are results [38], and diamonds are our present result. The curves show the TALYS1.95 computations.*

### CONCLUSIONS

In experiments on the electron linac LUE-40 of RDC "Accelerator" NSC KIPT the flux-averaged cross-sections $\langle σ(E_{γmax}) \rangle$ of photonuclear reactions $^{100}$Mo(γ,n)$^{99}$Mo, $^{27}$Al(γ,x)$^{24}$Na, $^{93}$Nb(γ,n)$^{92m}$Nb, $^{93}$Nb(γ,3n)$^{90}$Nb, and $^{181}$Ta(γ,n)$^{180g}$Ta were measured using the γ-activation technique in the bremsstrahlung end-point energy range $E_{γmax} = 35–95$ MeV. These results were obtained in two different experimental approaches (Setup 1 and 2) for cleaning a beam of bremsstrahlung gamma rays from the impurity of the electronic component. A comparison of the experimental flux-average cross-sections measured for Setup 1 and Setup 2 was performed and showed good agreement.

The theoretical flux-average cross-sections $\langle\sigma(E_{\gamma max})\rangle_{th}$ was computed using the partial cross-section values $\sigma(E)$ with the TALYS1.9-1.95 codes and bremsstrahlung γ-flux calculated by GEANT4.9.2.

It is shown that the normalization coefficients $k_{monitor}$, which display the deviation of the calculated in the GEANT4 code bremsstrahlung flux from the real flux that fell on the target, are the same for investigated reactions.

The possibility of using the $^{100}$Mo(γ,n)$^{99}$Mo, $^{27}$Al(γ,x)$^{24}$Na, $^{93}$Nb(γ,n)$^{92m}$Nb, $^{93}$Nb(γ,3n)$^{90}$Nb, and $^{181}$Ta(γ,n)$^{180g}$Ta reactions for monitoring of the bremsstrahlung flux of gamma quanta in the end-point energy range of 30-100 MeV is considered.

It has been established that molybdenum, aluminum and niobium can be used as monitors of the bremsstrahlung quantum flux. Mo and Nb have a large cross-section of monitor reactions. Hence, higher statistics of measurements and less statistical error. Also, Mo and Nb are convenient, because cover the energy range from 10 MeV. It is important, that the theoretical average cross-section from the Talys1.9-1.95 code coincides with the experimental one for reactions $^{100}$Mo(γ,n)$^{99}$Mo.

Tantalum is difficult to use as a monitor for the $^{181}$Ta(γ,n)$^{180g}$Ta using reaction due to a large tabular error and very complex γ-line picture of induced activity spectrum.

Aluminum is the most affordable material for flux monitoring. This reaction has a small cross-section and a high reaction threshold. At the same time, the emission spectrum of $^{24}$Na contains a small number of high-intensity lines, and the cross-section of $^{27}$Al(γ,x)$^{24}$Na is well investigated in literature [3,28]. Therefore, using this reaction as a monitor of gamma-flux is the most reliable bremsstrahlung end-point energy range $E_{\gamma max} > 45$ MeV.